\documentclass{jetpl}
\twocolumn

\lat

\title{Giant enhancement of quantum decoherence by frustrated environments}

\rtitle{Giant enhancement\ldots}

\sodtitle{Giant enhancement of quantum decoherence by frustrated environments}

\author{S.\, Yuan$^*$,
M.\, I.\, Katsnelson$^+$\/\thanks{e-mail: M.Katsnelson@science.ru.nl}, and H.\, De\, Raedt$^*$}

\rauthor{S.\, Yuan, M.\, I.\, Katsnelson, and H.\, De\, Raedt}

\sodauthor{Yuan, Katsnelson, De Raedt}

\address{$^*$Department of Applied Physics, Materials Science Center,
University of Groningen, Nijenborgh 4, NL-9747 AG Groningen, The Netherlands\\~\\
$^+$Institute of Molecules and Materials, Radboud
University of Nijmegen, NL-6525 ED Nijmegen, The Netherlands}

\dates{7 April 2006}{*}

\abstract{This Letter studies the decoherence in a system of two antiferromagnetically
coupled spins that interact with a spin bath environment.
Systems are considered that range from the rotationally invariant
to highly anisotropic spin models, have different topologies
and values of parameters that are fixed or are allowed to fluctuate randomly.
We explore the conditions under which the two-spin system clearly
shows an evolution from the initial spin-up - spin-down
state towards the maximally entangled singlet state.
We demonstrate that frustration and, especially, glassiness of the spin environment
strongly enhances the decoherence of the two-spin system.}

\PACS{03.65.Yz,75.10.Nr}

\begin{document}

\maketitle

The interaction between a quantum system, called central system in
what follows, and its environment affects the state of the former.
Intuitively, we expect that by turning on the interaction with the
environment, the fluctuations in the environment will lead to a
reduction of the coherence in the central system. % and its entropy growth.
This process is called decoherence~\cite{zeh,zurek}. In
general, there are two different mechanisms that contribute to
decoherence. If the environment is dissipative (or coupled to a
dissipative system), the total energy is not conserved and the
central system + environment relax to a stationary equilibrium
state, for instance the thermal equilibrium state. In this paper,
we exclude this class of dissipative processes and restrict
ourselves to closed quantum systems in which a small, central
system is brought in contact with a larger quantum system that is
prepared in its ground state. Then, decoherence is solely due to
fact that the initial product state (wave function of the central
system times wave function of the environment) evolves into an
entangled state of the whole system. The interaction with the
environment causes the initial pure state of the central system to
evolve into a mixed state, described by a reduced density
matrix\cite{neumann}, obtained by tracing out all the degrees of
freedom of the environment~\cite{zeh,zurek,feynman,leggett}.

Not all initial states are equally sensitive to decoherence. The
class of states that is ``robust'' with respect to the interaction
with the environment are called pointer states~\cite{zurek}. If
the Hamiltonian of the central system is a perturbation, relative
to the interaction Hamiltonian $H_{int}$, the pointer states are
eigenstates of $H_{int}$~\cite{zurek,paz}. In this case, the
pointer states are essentially ``classical states'', such as
states with definite particle positions or with definite spin
directions of individual particles for magnetic systems. In
general, these states, being a product
of states of individual particles forming the system, are not entangled.
On the other hand, decoherence does not necessarily imply that the central
system evolves to a classical-like state. If $H_{int}$ is much
smaller than the typical energy differences in the central system,
the pointer states are eigenstates of the latter, that is, they
may be ``quantum'' states such as standing waves, stationary
electron states in atoms, tunnelling-split states for a particle
distributed between several potential wells, singlet or triplet
states for magnetic systems, etc.~\cite{paz}. This may explain,
for example, that one can observe linear atomic spectra - the
initial states of an atom under the equilibrium conditions are
eigenstates of its Hamiltonian and not arbitrary superpositions
thereof.

Let us now consider a central system for which the ground state
is a maximally entangled state, such as a singlet.
In the absence of dissipation and for an environment that is in the ground state before
we bring it in contact with this central system, the loss
of phase coherence induces one of following
qualitatively different types of behavior:
\begin{enumerate}
\item{The interaction/bath dynamics is such that there is very little relaxation.}
\item{The system as a whole relaxes to some state (which may or may not be close to the ground state)
and this state is a complicated superposition of the states of the central system and the environment.}
\item{The system as a whole relaxes to a state that is (to good approximation)
a direct product of the states of the central system and a superposition of states of the environment.
In this case there are two possibilities:}
\begin{enumerate}
\item{The central system does not relax to its ground state;}
\item{The central system relaxes to its maximally entangled ground state.}
\end{enumerate}
\end{enumerate}
Only case 3b is special:
The environment and central system are not entangled
(to a good approximation) but nevertheless the decoherence induces a
very strong entanglement in the central system.
In this paper, we demonstrate
that, under suitable conditions, dissipation free decoherence
forces the central system to relax to a maximally entangled state
which itself, shows very little entanglement with the state of the environment.

Most theoretical investigations of decoherence have been carried out for
oscillator models of the environment for which powerful
path-integral techniques can be used to treat the environment
analytically~\cite{feynman,leggett}. On the other hand, it has
been pointed out that a magnetic environment, described by quantum
spins, is essentially different from the oscillator model in many
aspects~\cite{stamp}. For the simplest model of a single spin in
an external magnetic field, some analytical results are
known~\cite{stamp}. For the generic case of two and more spins,
numerical simulation~\cite{ourPRL,ourPRE} is the main source of
theoretical information.
Not much is known now about which physical properties of the
environment are important for the efficient selection of pointer
states. Recent numerical simulations~\cite{ourPRE} confirm the
hypothesis~\cite{zurek_nature} on the relevance of the chaoticity
of the environment but its effect is actually not drastic.

\begin{table*}[tb]
\caption{Table I. Minimum value of the correlation of the central spins
and the energy of the whole system (which is conserved), as observed during the time evolution
corresponding to the curves listed in the first column.
The correlations $\langle {\bf S}_1\cdot {\bf S}_2\rangle_0$ and the ground state
energy $E_0$ of the whole system are obtained by numerical diagonalization
of the Hamiltonian Eq.(\ref{HAM}).
}
\begin{center}
\begin{tabular}{l|cc|cc}
\noalign{\vskip 2pt}
%\hline
&
$\langle \Psi(t)| H |\Psi(t)\rangle$ & $E_0$&
$\min_{t}\langle {\bf S}_1(t)\cdot {\bf S}_2(t)\rangle$ & $\langle {\bf S}_1\cdot {\bf S}_2\rangle_0$
  \\
\hline
%\noalign{\vskip 4pt}
 Fig.~1 (a)  &   -1.299  & -1.829 &  -0.659  &  -0.723   \\
 Fig.~1 (b)  &   -1.532  & -2.065 &  -0.695  &  -0.721   \\
 Fig.~1 (c)  &   -1.856  & -2.407 &  -0.689  &  -0.696   \\
 Fig.~2      &   -4.125  & -4.627 &  -0.744  &  -0.749   \\
 Fig.~3 (a)  &   -1.490  & -1.992 &  -0.746  &  -0.749   \\
 Fig.~3 (b)  &   -0.870  & -1.379 &  -0.260  &  -0.741   \\
 Fig.~3 (c)  &   -1.490  & -1.997 &  -0.737  &  -0.744   \\
 Fig.~3 (d)  &   -2.654  & -3.160 &  -0.742  &  -0.745   \\
 Fig.~3 (e)  &   -7.791  & -8.293 &  -0.716  &  -0.749   \\
 Fig.~3 (f)  &   -3.257  & -3.803 &  -0.713  &  -0.718   \\
 Fig.~4 (b)  &   -0.884  & -1.388 &  -0.424  &  -0.733   \\
 Fig.~4 (c)  &   -1.299  & -1.829 &  -0.659  &  -0.723   \\
 Fig.~4 (d)  &   -1.299  & -1.807 &  -0.741  &  -0.743   \\
 Fig.~4 (e)  &   -1.843  & -2.365 &  -0.738  &  -0.735   \\
%\hline
\end{tabular}
\label{table}
\end{center}
\end{table*}

In this paper, we report on the results of numerical simulations of quantum spin
systems, demonstrating the crucial role of frustrations in the
environment on decoherence.
In particular, we show that, under appropriate conditions,
decoherence can cause an initially classical state of the central system
to evolve into the most extreme, maximally entangled state.
We emphasize that we only consider systems in which
the total energy is conserved such that the decoherence is
not due to dissipation.

We study a model in which
two antiferromagnetically coupled spins, called the central system,
interact with an environment of spins. The model
is defined by
\begin{eqnarray}
H&=&H_c+H_e+H_{int}
\quad,\quad
%\nonumber\\
H_c=-J\mathbf{S}_{1}\cdot\mathbf{S}_{2}
\nonumber\\
H_e&=&-\sum_{i=1}^{N-1}\sum_{j=i+1}^{N}\sum_{\alpha} \Omega_{i,j}^{(\alpha)}I_{i}^{\alpha}I_{j}^{\alpha}
\nonumber\\
%\quad,\quad
H_{int}&=&-\sum_{i=1}^{2}\sum_{j=1}^{N}\sum_{\alpha} \Delta_{i,j}^{(\alpha)}S_{i}^{\alpha}I_{j}^{\alpha}
, \label{HAM}%
\end{eqnarray}
where the exchange integrals $J<0$ and $\Omega_{i,j}^{(\alpha)}$
determine the strength of the interaction between spins
$\mathbf{S}_n=\left(  S^{x}_n,S^{y}_n,S^{z}_n\right)$ in the
central system ($H_c$), and the spins $\mathbf{I}_n=\left(
I^{x}_n,I^{y}_n,I^{z}_n\right)$ in the environment ($H_e$),
respectively. The exchange integrals $\Delta_{i,j}^{(\alpha)}$
control the interaction ($H_{int}$) of the central system with its
environment. In Eq.~(\ref{HAM}), the sum over $\alpha$
runs over the $x$, $y$ and $z$ components of spin $1/2$ operators.
The number of spins in the environment is $N$.

Initially, the central system is in the spin-up - spin-down state and
the environment is in its ground state.
Thus, we write the initial state as
$|\Psi(t=0)\rangle=|\uparrow\downarrow\rangle|\Phi_0\rangle$.
The time evolution of the system is obtained by
solving the time-dependent Schr\"odinger equation for the
many-body wave function $|\Psi(t)\rangle$, describing the central system
plus the environment. The numerical method that we use is described in
Ref.~\cite{method}. It conserves the energy of the whole system
to machine precision.

\begin{figure*}[t]
\begin{center}
\mbox{
\includegraphics[width=8.0cm]{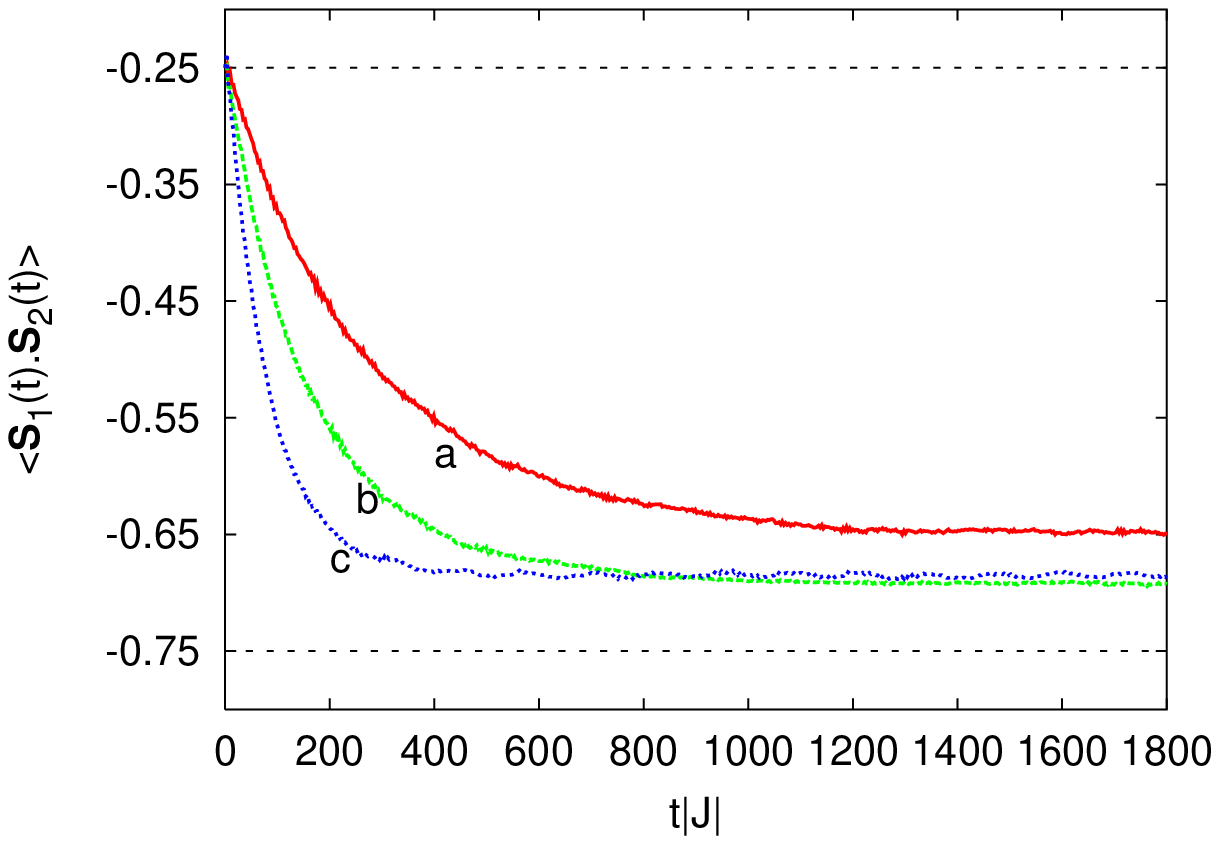}
\includegraphics[width=8.0cm]{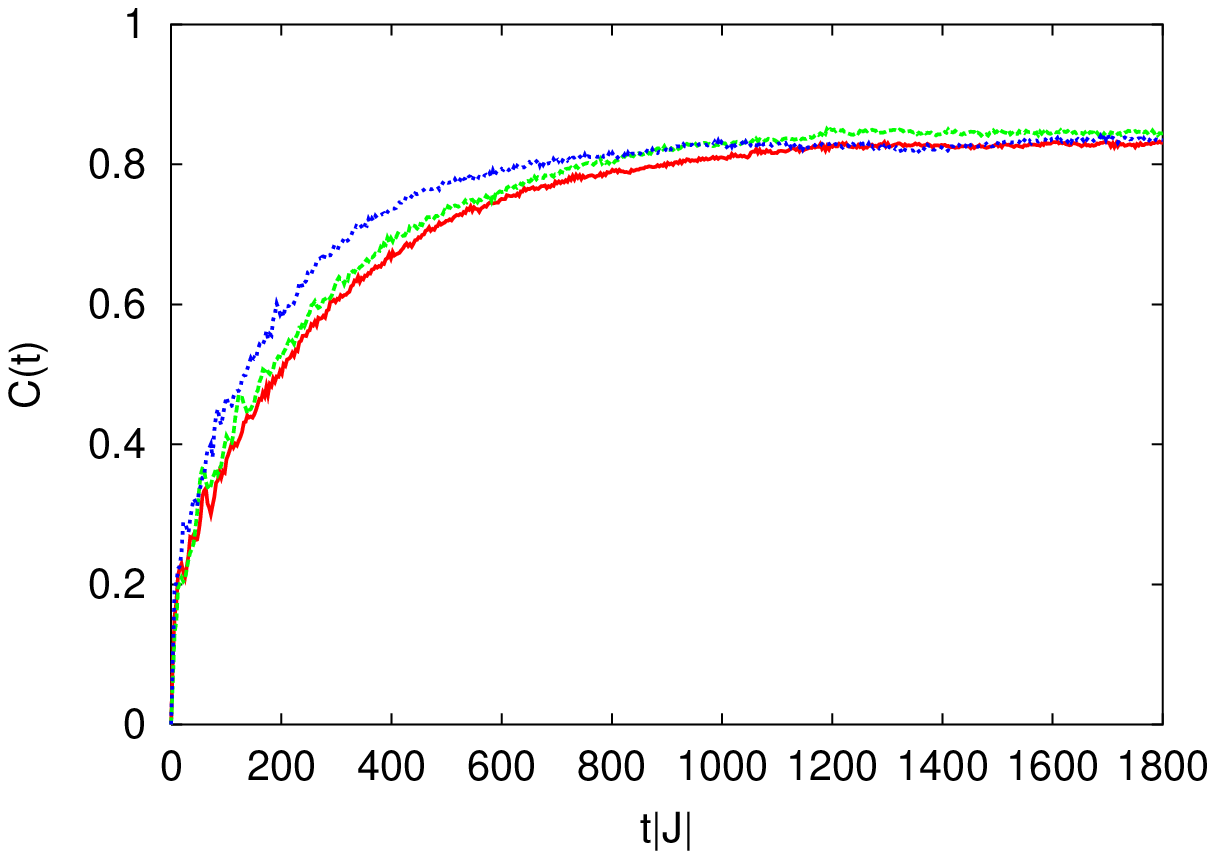}
}
\caption{Fig.~1 (color online)
Left:
Time evolution of the correlation
$\langle \Psi(t)| {\bf S}_1\cdot {\bf S}_2|\Psi(t)\rangle$
of the two spins in the central system.
Dashed horizontal line at -1/4: Correlation in the
initial state ($\langle \Psi(t=0)| {\bf S}_1\cdot {\bf S}_2|\Psi(t=0)\rangle=-1/4$);
Horizontal line at -3/4: Expectation value in the singlet state;
(a) Environment containing $N=14$ quantum spins;
(b) $N=16$;
(c) $N=18$.
The parameters $\Omega_{i,j}^{(\alpha)}$ and $\Delta_{i,j}^{(\alpha)}$
are uniform random numbers in the range $[-0.15|J|,0.15|J|]$.
Right:
Time evolution of the concurrence $C(t)$ for three different
random realizations of a spin glass environment. The parameters
are uniform random numbers in the range $-0.15|J|\le
\Omega_{i,j}^{(\alpha)}, \Delta_{i,j}^{(\alpha)} \le0.15|J| $ and
the environment contains $N=14$ quantum spins. The transition from
an unentangled state ($C(t)=0$) to a nearly fully entangled state
($C(t)=1$) is clearly seen.
}
\label{C12}%
\end{center}
\end{figure*}

\begin{figure}[t]
\begin{center}
\includegraphics[width=8cm]{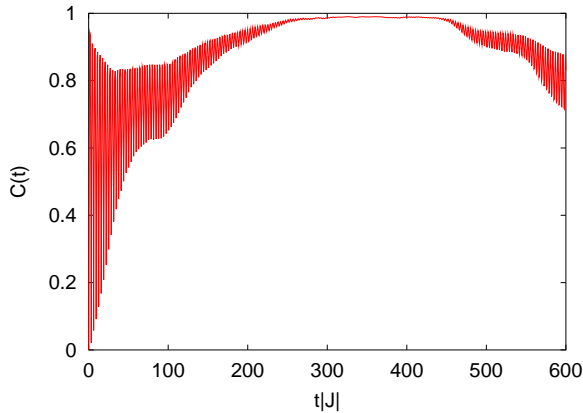}
\caption{Fig.~2 (color online)
Time evolution of the concurrence $C(t)$
for the case of a frustrated antiferromagnetic environment.
The interactions of the central system
and the environment are uniform random numbers in the range
$-0.15|J|\le \Delta_{i,j}^{(\alpha)} \le-0.05|J|$.
The environment contains $14$ quantum spins,
arranged on a triangular lattice and interacting
with nearest neighbors only.
The nonzero exchange integrals are uniform random numbers in the range
$-0.55|J|\le \Omega_{i,j}^{(\alpha)} \le-0.45|J| $.
The transition from an unentangled state ($C(t)=0$) to a
nearly fully entangled state ($C(t)=1$) is evident,
as is the onset of recurrent behavior due to the finite
size of the environment.
}
\label{TRI}%
\end{center}
\end{figure}

%\begin{figure}[t]
%\begin{center}
%\includegraphics[width=8cm]{figure1.eps}
%\caption{(color online)
%Time evolution of the correlation
%$\langle \Psi(t)| {\bf S}_1\cdot {\bf S}_2|\Psi(t)\rangle$
%of the two spins in the central system.
%Dashed horizontal line at -1/4: Correlation in the
%initial state ($\langle \Psi(t=0)| {\bf S}_1\cdot {\bf S}_2|\Psi(t=0)\rangle=-1/4$);
%Horizontal line at -3/4: Expectation value in the singlet state;
%(a) Environment containing $N=14$ quantum spins;
%(b) $N=16$;
%(c) $N=18$.
%The parameters $\Omega_{i,j}^{(\alpha)}$ and $\Delta_{i,j}^{(\alpha)}$
%are uniform random numbers in the range $[-0.15|J|,0.15|J|]$.
%}
%\label{C12}%
%\end{center}
%\end{figure}
%
%\begin{figure}[t]
%\begin{center}
%\includegraphics[width=8.5cm]{figure2.eps}
%\caption{(color online)
%Time evolution of the concurrence $C(t)$ for three different
%random realizations of a spin glass environment. The parameters
%are uniform random numbers in the range $-0.15|J|\le
%\Omega_{i,j}^{(\alpha)}, \Delta_{i,j}^{(\alpha)} \le0.15|J| $ and
%the environment contains $N=14$ quantum spins. The transition from
%an unentangled state ($C(t)=0$) to a nearly fully entangled state
%($C(t)=1$) is clearly seen. }
%\label{CONC}%
%\end{center}
%\end{figure}

\begin{figure}[t]
\begin{center}
\includegraphics[width=8cm]{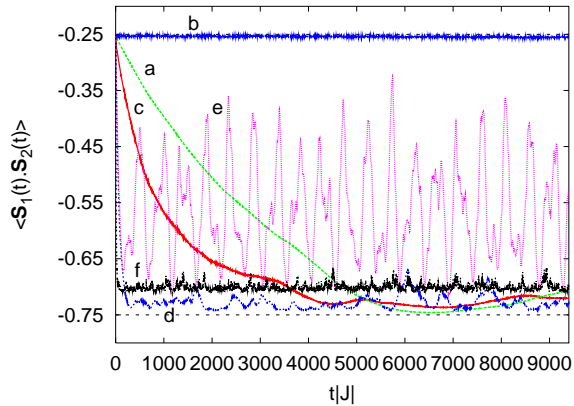}
\caption{Fig.~3 (color online) Time evolution of the correlation $\langle \Psi (t)|%
\mathbf{S}_{1}\cdot \mathbf{S}_{2}|\Psi (t)\rangle $ of the two spins in the
central system. Environment containing $N=16$ quantum spins. Dashed
horizontal line at -1/4: Correlation in the initial state ($\langle \Psi
(t=0)|\mathbf{S}_{1}\cdot \mathbf{S}_{2}|\Psi (t=0)\rangle =-1/4$);
Horizontal line at -3/4: Expectation value in the singlet state.
For all curves (a-f) $\Delta _{i,j}^{(x)}=\Delta _{i,j}^{(y)}=0$, that is $H_{int}$
is Ising like.
The values of $\Delta _{i,j}^{(z)}$ are: (a)
random $-0.0375\left\vert J\right\vert $ or $0.0375\left\vert J\right\vert $,
(b-e) random $-0.075\left\vert J\right\vert $ or $0.075\left\vert
J\right\vert $, (f) random $-0.15\left\vert J\right\vert $ or $%
0.15\left\vert J\right\vert $.
The values of $\Omega_{i,j}^{(\alpha )}$ are uniform
random numbers in the range: (b) $[-0.0375|J|,0.0375|J|]$, (a,c) $%
[-0.15|J|,0.15|J|]$, (d,f) $[-0.3|J|,0.3|J|]$ and (e) $[-|J|,|J|]$.
}
\label{FIG6}%
\end{center}
\end{figure}

\begin{figure}[t]
\begin{center}
\includegraphics[width=8cm]{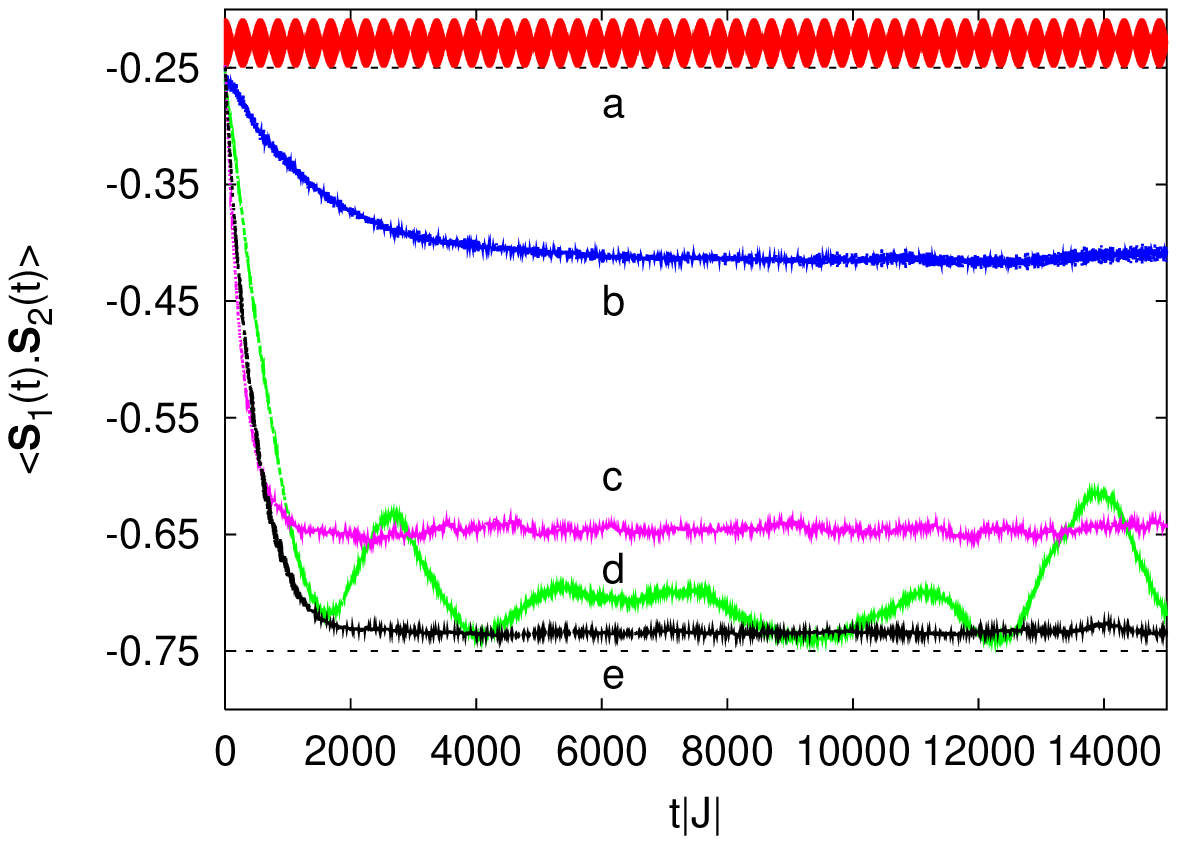}
\caption{Fig.~4 (color online)
Effect of the symmetry of the
exchange interactions $\Omega_{i,j}^{(\alpha)}$ and $\Delta_{i,j}^{(\alpha)}$
on the time evolution of the correlation
$\langle \Psi(t)| {\bf S}_1\cdot {\bf S}_2|\Psi(t)\rangle$
of the two spins in the central system.
Dashed horizontal line at -1/4: Correlation in the
initial state ($\langle \Psi(t=0)| {\bf S}_1\cdot {\bf S}_2|\Psi(t=0)\rangle=-1/4$);
Horizontal line at -3/4: Correlation in the singlet state;
Other lines from top to bottom (at $t|J|=6000$):
(a) Ising $H_{int}$ with Ising $H_{e}$, $N=14$;
(b) Heisenberg-like $H_{int}$ with Ising $H_{e}$, $N=14$;
(c) Heisenberg-like $H_{int}$ with Heisenberg-like $H_{e}$, $N=14$;
(d) Ising $H_{int}$ with Heisenberg-like $H_{e}$, $N=14$;
(e) Same as (d) except that $N=18$.
We use the term Heisenberg-like $H_{int}$ ($H_{e}$) to indicate
that $\Delta_{i,j}^{(\alpha)}$ ($\Omega_{i,j}^{(\alpha)}$)
are uniform random numbers in the range
$[-0.15\left\vert J\right\vert ,0.15\left\vert J\right\vert ]$.
Likewise, Ising $H_{int}$ ($H_{e}$) means
that $\Delta_{i,j}^{(x,y)}=0$ ($\Omega_{i,j}^{(x,y)}=0$), and $\Delta
_{i,j}^{(z)}$ ($\Omega_{i,j}^{(z)}$) are random $-0.075\left\vert J\right\vert
$ or $0.075\left\vert J\right\vert $.
}
\label{FIG59}%
\end{center}
\end{figure}

By changing the parameters of model (\ref{HAM}), we explore the
conditions under which the central system clearly shows an
evolution from the initial spin-up - spin-down state towards the
maximally entangled singlet state. We consider systems that range
from the rotationally invariant Heisenberg case to the extreme
case in which $H_e$ and $H_{int}$ reduce to the Ising model,
topologies for which the central system couples to two and to all
spins of the environment, and values of parameters that are fixed
or are allowed to fluctuate randomly. Illustrative results of
these calculations are shown in Figs.~1 - 4.
In Table I, we present the corresponding numerical data
of the energy $\langle \Psi(0)| H |\Psi(0)\rangle=\langle \Psi(t)| H |\Psi(t)\rangle$)
and of the two-spin correlation
$\langle {\bf S}_1(t)\cdot {\bf S}_2(t)\rangle=\langle \Psi(t)| {\bf S}_1\cdot {\bf S}_2|\Psi(t)\rangle$.
For comparison, Table I also contains the results of the
energy $E_0$ and of the two-spin correlation
$\langle {\bf S}_1\cdot {\bf S}_2\rangle_0$.
in the ground state of the whole system,
as obtained by numerical diagonalization of the Hamiltonian Eq.(\ref{HAM}).

We monitor the effects of decoherence by computing the
expectation value $\langle \Psi(t)| {\bf S}_1\cdot {\bf S}_2|\Psi(t)\rangle$.
The central system is in the singlet state if $\langle {\bf
S}_1(t)\cdot {\bf S}_2(t)\rangle=-3/4$, that is if $\langle {\bf
S}_1(t)\cdot {\bf S}_2(t)\rangle$ reaches its minimum value. We
also study the time evolution of the concurrence $C(t)$, which is
a convenient measure for the entanglement of the spins in the
central system~\cite{Wootters98}. The concurrence is equal to one
if the central system is in the singlet state and is zero for an
unentangled pure state such as the spin-up - spin-down
state~\cite{Wootters98}.

A very extensive search
through parameter space leads to the following conclusions:
\begin{itemize}
\item{The maximum amount of entanglement
strongly depends on the values of the model parameters
$\Omega_{i,j}^{(\alpha)}$ and $\Delta_{i,j}^{(\alpha)}$.
For the case in which there is strong decoherence,
increasing the size of the environment will enhance the decoherence
in the central system (compare the curves of Fig.~1(a,b,c) and Fig.~4(d,e)).
Keeping the size of the environment fixed,
different realizations of the random parameters do not significantly
change the results for the correlation and concurrence (right panel of Fig.~1).
However, the range of random values
$\Omega_{i,j}^{(\alpha)}$ and $\Delta_{i,j}^{(\alpha)}$
for which maximal entanglement can be
achieved is narrow, as illustrated in Figs.~3 and 4.
In Fig.~3 we compare results
for the same type of  $H_{int}$ (Ising like) and
the same type of $H_e$ (anisotropic Heisenberg like),
but with different values of the model parameters.
In Fig.~4, we present results
for different types of $H_{int}$ and $H_e$.
but for parameters within the same range.}
\item{Environments that exhibit some form of frustration, such as
spin glasses or frustrated antiferromagnets, may be very effective
in producing a high degree of entanglement between the two central
spins, see Figs.~1-4.}
\item{Decoherence is most effective if the exchange couplings
between the system and the environment are random (in a limited range)
and anisotropic, see Figs.~3 and 4.}
\item{The details of the internal dynamics of the environment
affects the maximum amount of entanglement that can be achieved~\cite{ourPRE},
and also affects the speed of the initial relaxation 
(compare the curves of Fig.~3(b,c,d,e), Fig.~4(a,d) and Fig.~4(b,c)).}
\item{For the case in which there is strong decoherence,
for the same $H_e$ and the same type of $H_{int}$, decreasing the strengh of $H_{int}$
will reduce the relaxation to the finial state, 
and the final state comes closer to the singlet state 
(compare the curves of Fig.~3(a,c) and Fig.~3(d,f)).}
\end{itemize}

% Our data analysis shows that for all cases under consideration,
% the typical speed of the initial decay of the correlation
% is proportional to $\sum_{ij\alpha}
% |\Delta_{i,j}^{(\alpha)}|$ and that this estimate
% does not depend on the model in an essential manner.
% At the same time, the efficiency of the environment,
% as measured by the amplitude of decaying ``Rabi oscillations'',
% strongly depends on the character of the spin-spin
% interactions in the environment.
% From our results, it follows that decoherence is very efficient
% if the environment is frustrated and that in this case, adding a
% few number of spins to the environment leads to a strong
% suppression of finite-size fluctuations. An example of this effect
% is shown in Fig.~\ref{FIG59} for the case of $N=14$ (green line)
% and $N=18$ (black line) spins in the glassy environment.

Earlier simulations for the Ising model in a transverse field
have shown that time-averaged distributions of the energies
of the central system and environment agree with those
of the canonical ensemble at some effective temperature~\cite{jens85,sait96}.
Our results do not contradict these findings but show that
there are cases in which the central system relaxes from a
high energy state to its ground state while the environment
starts in the ground state and ends up in state
with slightly higher energy. As shown in Fig.4(e), this
state is extremely robust and shows very little fluctuations.

For the models under consideration, the efficiency of decoherence decreases drastically in
the following order: Spin glass (random long-range interactions of both
signs); Frustrated antiferromagnet (triangular lattice with the
nearest-neighbour interactions); Bipartite antiferromagnet (square
lattice with the nearest-neighbour interactions); One-dimensional
ring with the nearest-neighbour antiferromagnetic interactions.
This can be understood as follows.
A change of the state of the central system affects a group of spins
in the environment. The suppression of
off-diagonal elements of the reduced density matrix
can be much more effective if the group of disturbed spins is larger.
The state of the central system is the most flexible
in the case of a coupling to a spin glass for which, in the thermodynamic
limit, an infinite number of infinitely closed quasi-equilibrium
configurations exist \cite{SG,parisi}. As a result, a very small
perturbation leads to the change of the system {\it as a whole}.
This may be considered as a quantum analog of the phenomenon of
``structural relaxation'' in glasses. This suggests that frustrated
spin systems that are close to the glassy state should provide
extremely efficient decoherence.

%We conjecture that this observation may be relevant for a broad range of physical
%systems, including high-temperature superconductors. In the
%underdoped regime, they demonstrate some clear evidence of
%competing spin interactions and glassy behaviour
%\cite{chou,julien,keimer}. Qualitatively, one may assume that this
%should lead to unusually strong decoherence of the Cooper pair
%condensate and hence to suppression of the superconductivity. The
%famous pseudogap phenomenon in the underdoped high-temperature
%superconductors \cite{pseudo1} is explained sometimes in terms of
%``remnant'' Cooper pairing without macroscopic coherence and
%long-range off-diagonal order \cite{pseudo2,pseudo3}. In this
%physical picture, the role of QD due to the spin-glass environment
%is worth further investigation.
%It is also worthwhile to note that the dipole-dipole interactions
%of nuclear spins are long-ranged and can be of arbitrary sign. Our
%results then suggest that the nuclear spin thermal bath, being a
%highly frustrated system and abundantly present in
%magnetic materials, may be a more efficient decohering
%factor than anticipated before.

To conclude, we have demonstrated that frustrations and,
especially, glassiness of the spin environment result in a very
strong enhancement of its decohering action on the central spin
system.
Our results convincingly show that this enhancement can be so strong that
solely due to decoherence,
a fully disentangled state may evolve into a fully entangled state,
even if the environment contains a relatively small numbers of spins.

\end{document}